# Comparative study of autodyne and heterodyne laser interferometry for imaging

Eric Lacot,[*] Olivier Jacquin, Grégoire Roussely, Olivier Hugon,

and Hugues Guillet de Chatellus

*Laboratoire de Spectrométrie Physique, Unité Mixte de Recherche 5588, Université Joseph*

*Fourier de Grenoble, Centre National de la Recherche Scientifique,*

*B.P. 87 38402 Saint Martin d'Hères, France*

[*]*Corresponding author: eric.lacot@ujf-grenoble.fr*

For given laser output power, object under investigation and photodiode noise level, we have theoretically compared the signal to noise ratio (SNR) of a heterodyne scanning imager based on a Michelson interferometer and of an autodyne setup based on the Laser Optical Feedback Imaging (LOFI) technique. In both cases, the image is obtained point by point. In the heterodyne configuration, the beating between the reference beam and the signal beam is realized outside the laser cavity (i.e. directly on the detector), while in the autodyne configuration, the wave beating takes place inside the laser cavity and therefore is indirectly detected. In the autodyne configuration, where the laser relaxation oscillations play a leading role, we have compared 1D scans obtained by numerical simulations with different lasers dynamical parameters. Finally we have determined the best laser for LOFI applications and the experimental conditions for which the LOFI detection setup (autodyne interferometer) is competitive comparing to a heterodyne interferometer. © *2010 Optical Society of America*

## 1. INTRODUCTION

Photodetectors are sensitive to the power (photon flux) but not to the phase of an incident optical wave. However, the complex amplitude (magnitude and phase) of this wave can be measured by mixing it with a coherent reference wave of stable frequency. By shifting the frequency of one of these waves, we get the so-called heterodyne interferometry. Resulting from this shift, the interference between the two waves produces an intensity modulation at the beat frequency measured by the photodetector. In this paper, we refer to autodyne laser interferometry when the heterodyne wave mixing takes place inside the cavity of the laser source, while we speak about heterodyne laser interferometry when the mixing is realized directly on the photodetector (i.e. outside the laser cavity).

Since the development of the first laser in 1960, laser heterodyne interferometry has become a useful technique at the basis of high accuracy measurement systems for scientific and industrial applications [1]. For imaging purposes, optical heterodyning in combination with laser scanning microscopy was suggested more than 30 years ago [2] and was later combined with low coherence light for biological or medical observation [3-6]. Heterodyne interferometry is photon shot-noise limited and therefore extremely sensitive [7-9].

More recently, the sensitivity of laser dynamics to frequency-shifted optical feedback has been used in autodyne interferometry, for example in self-mixing laser Doppler velocimetry [10,11], vibrometry [12], near field microscopy [13,14] and laser optical feedback imaging (LOFI) experiments. [15-17]. More precisely, the laser optical feedback imaging (LOFI) technique is a powerful imaging method combining the great accuracy of optical interferometry with the very high sensitivity of class B lasers to optical feedback [18,19]. In this autodyne method, thanks to a resonant amplification of the optical beating, relative feedback power as low as $10^{-13}$ (-130 dB) are then detectable in a 1 KHz detection bandwidth, with a laser output power of a few milliwatts only. The LOFI method is therefore also shot noise limited [18].

Compared to conventional optical heterodyne detection, frequency-shifted optical feedback shows intensity modulation contrast higher by several orders of magnitude (typically $10^3$ for a laser diode and $10^6$ for a microchip laser) [15]. The maximum of the modulation is obtained when the shift frequency is resonant with the laser relaxation oscillation frequency, typically 1 GHz for a laser diode and 1 MHz for a microchip laser. The signal gain is given by the cavity damping rate to the population-inversion damping rate ratio [15,18]. In the resonant condition, an optical feedback level as low as -170 dB (i.e. $10^{-17}$ compared to the intracavity power) has been detected [11]. In a LOFI experiment, the beating signal and the laser quantum noise are both resonantly amplified near the relaxation frequency, i.e. they follow the same gain curves. As a result, the signal to noise ratio of the LOFI method is frequency independent [18]. Therefore, the laser relaxation frequency and the value of the LOFI gain seem to have no particular importance.

The objective of this paper is twofold: firstly to determine the best laser to realize an autodyne LOFI interferometer and secondly to determine analytically the real gain and the main advantages of an autodyne LOFI interferometer compared to a conventional heterodyne Michelson interferometer. This paper is organized as follows. In the first section, we give a basic description of the two types (autodyne and heterodyne) of experimental setups. In the second section, we recall the LOFI signal induced by the resonant beating inside the laser cavity, the LOFI noise induced by the laser quantum noise and finally, for a given detection noise level, we determine the laser parameters which optimize (sensitivity, dynamic range, ..) the LOF imaging setup for reflectivity measurements. In the third section, we compare the signal to noise ratio of the Michelson setup (heterodyne detection) to the LOFI setup (autodyne detection), to determine the real gain of the LOFI method. Finally we conclude clearly on the advantages and disadvantages of the LOFI method compared to a Michelson interferometer in terms of signal to noise ratio, dynamic range and simplicity of implementation.

## 2. STUDIED INTERFEROMETRIC SETUPS

A schematic diagram of the LOFI experimental setup (i.e. the autodyne experimental interferometer) is shown in Fig. 1(a). Typically, the laser is a CW $Nd^{3+}$:YAG microchip lasing at the wavelength λ=1064 nm with an output power of the order of $P_{out} \approx 10\,mW$ and a typical relaxation frequency of $F_R \approx 1\,MHz$ [20]. The laser beam is sent towards a frequency shifter that is composed of two acousto-optic deflectors (AOD), respectively supplied by a RF signal at 81.5 MHz and 81.5 MHz+$F_e$/2 where $F_e$ is the shift frequency after a round trip. The diffracted beam (order -1) of the first AOD and the diffracted beam (order+1) of the second AOD are selected. At this stage, the resulting optical frequency shift of the laser beam is $F_e$/2. Next, the beam is sent to the target using a galvanometric scanner composed by two rotating mirrors. A part of the light diffracted and/or scattered by the target is then reinjected inside the laser cavity after a second pass in the galvanometric scanner and the frequency shifter. Therefore, the optical frequency of the reinjected light is shifted by $F_e$. This frequency can be adjusted and is typically of the order of the laser relaxation frequency $F_R$. For the geometrical point of view, the laser beam waist and the laser focal spot on the target under investigation are optically conjugated through the Lenses $L_1$ and $L_2$. The amount of optical feedback is characterized by the effective power reflectivity ($R_e$) of the target, where $R_e$ takes into account, the target albedo, the numerical aperture of the collecting optics, the AOD efficiencies and the transmission of all of the optical components (except for the beam splitter which is addressed separately) and the overlap of the retro-diffused field with the Gaussian cavity beam [18]. In the case of a weak optical feedback, the coherent interaction (beating) between the lasing electric field and the frequency-shifted reinjected field leads to a modulation of the laser output power. For the detection purpose, a fraction of the output beam of the microchip laser is sent to a photodiode by means of a beam splitter characterized by a power reflectivity $R_{bs}$. The photodiode is assumed to have a quantum efficiency of 100%. The voltage delivered by the photodiode is finally analyzed by a lock-in amplifier which gives the LOFI signal (i.e. the magnitude and the phase of the retro-diffused electric field) at the demodulation frequency $F_e$ [21, 22]. Experimentally, the LOFI images are obtained pixel by pixel (i.e. point by point, line after line) by full 2D galvanometric scanning.

A schematic diagram of the Michelson experimental setup (i.e. the autodyne experimental interferometer) is shown on Fig. 1(b). Compared to Fig. 1(a), the only differences are: the optical isolator which prevents any optical feedback in the laser source, the beam splitter orientation and the reference mirror (RM) which allows the mixing of the reference wave with the signal wave directly on the detector. In this configuration, the delivered voltage is also analyzed by a lock-in amplifier.

Thus, for given laser output power, target and photodiode noise level, the experimental setups shown on Figs. 1(a) and 1(b), enable a direct comparison of the sensitivity of a heterodyne imaging setup based on a Michelson interferometer and of an autodyne interferometer based on a Laser Optical Feedback Imaging (LOFI) method. At this point, one can already notice that, compared to the heterodyne setup, the autodyne setup does not require complex alignment. More precisely, the LOFI setup is even always self-aligned because the laser simultaneously fulfills the function of the source (i.e. the emitter) and of the photodetector (i.e. the receptor).

## 3. AUTODYNE INTERFEROMETER (LOFI SETUP)

### A. LOFI signal

In the case of weak ($R_e \ll 1$) frequency shifted optical feedback, the dynamical behavior of a re-injected solid-state laser can be described by the following set of equations [18, 19]:

$$\frac{dN(t)}{dt} = \gamma_1 N_0 - \gamma_1 N(t) - BN(t)|E_c(t)|^2 + F_N(t), \qquad (1.a)$$

$$\frac{dE_c(t)}{dt} = \frac{1}{2}(BN(t) - \gamma_c)E_c(t) \\ + \gamma_e E_c(t-\tau_e)\cos\left[2\pi F_e t - 2\pi\left(\nu + \frac{F_e}{2}\right)\tau_e - \phi_c(t) + \phi_c(t-\tau_e)\right] + F_{E_c}(t) \qquad (1.b)$$

$$\frac{d\phi_c(t)}{dt} = 2\pi(\nu_c - \nu) + \gamma_e \frac{E_c(t-\tau_e)}{E_c(t)}\sin\left[2\pi F_e t - 2\pi\left(\nu + \frac{F_e}{2}\right)\tau_e - \phi_c(t) + \phi_c(t-\tau_e)\right] + F_{\Phi_c}(t). \quad (1.c)$$

where, N(t) is the population inversion, $E_c(t)$ and $\phi_c(t)$ are respectively the slowly varying amplitude and the optical phase of the laser electric field, $\nu_c$ is the laser cavity frequency which is assumed resonant with the atomic frequency, $\nu$ is the optical running laser frequency, B is the Einstein coefficient, $\gamma_1$ is the decay rate of the population inversion, $\gamma_c$ is the laser cavity decay rate, and $\gamma_1 N_0$ is the pumping rate. From a dynamical point of view, the optical feedback is characterized by three parameters, the optical shift frequency ($F_e$), the photon round trip time between the laser and the target ($\tau_e$) and the re-injection rate of the feedback electric field ($\gamma_e$) defined by:

$$\gamma_e = \gamma_c \sqrt{R_e} (1 - R_{bs}), \qquad (2)$$

where, compared to Ref. [18], we have taken the beam splitter reflectivity $R_{bs}$ into account to quantify the amount of light coming back inside the laser cavity [see Fig. 1]. Regarding the noise, the laser quantum fluctuations are described by the conventional Langevin noise functions $F_N(t)$, $F_{E_c}(t)$ and $F_{\Phi_c}(t)$, which have a zero mean value and a white noise type correlation function [23, 24]. In the set of Eqs. (1), the periodic functions express the beating (i.e. the coherent interaction) between the lasing and the feedback electric fields. The net laser gain is then modulated by the re-injected light at the optical shift frequency $F_e$. As a result, the photon output rate $p_{out}(t) = \gamma_c |E_c(t)|^2$ (number of photons per second) is periodically modulated [18]:

$$p_{out}(t) = \langle p_{out} \rangle + 2G(F_e) \langle p_{out} \rangle (1 - R_{bs}) \sqrt{R_e} \cos[2\pi F_e t + \varphi], \qquad (3)$$

where, $\langle p_{out} \rangle = \gamma_c \dfrac{\gamma_1}{B}(\eta - 1)$ is the mean value the photon output rate and $\eta = \dfrac{N_0}{\gamma_c / B}$ is the pumping parameter. Using Eq. (3), one can define (in the customary way for interferometry) the modulation contrast (MC) of the autodyne wave mixing:

$$\dfrac{\Delta p_{out}(F_e, R_e)}{\langle p_{out} \rangle} = 2G(F_e)(1 - R_{bs}) \sqrt{R_e}. \qquad (4)$$

In Eq. (4), $G(F_e)$ describes the amplification gain of the autodyne waves mixing, with:

$$G(F_e) = \frac{\gamma_c}{2\pi} \frac{\sqrt{\Delta F_R^2 + F_e^2}}{\sqrt{(F_R^2 - F_e^2)^2 + \Delta F_R^2 F_e^2}} \qquad (5)$$

where $F_R = \sqrt{\gamma_1 \gamma_c (\eta-1)}/2\pi$ is the laser relaxation oscillation frequency and $\Delta F_R = \gamma_1 \eta/2\pi$ is related to the damping rate of the relaxation oscillations [15].

At this point, one can noticed that Eqs. (3) (4) and (5) which have been obtained after a linearization of the set of Eqs. (1) are only valid for small modulation amplitude ($2G(F_e)(1-R_{bs})\sqrt{R_e} \ll 1$) and whatever the experimental conditions, the saturation of the laser output power modulation always limits the modulation contrast to unity ($0 \leq \frac{\Delta p_{out}(F_e, R_e)}{\langle p_{out} \rangle} \leq 1$)

Using a lock-in amplifier to demodulate the photodiode voltage, we finally obtain the LOFI signal, which we define as follows:

$$S_{LOFI}(R_e, F_e) = R_{bs} \frac{\Delta p_{out}(F_e, R_e)}{\sqrt{2}} = \frac{2G(F_e) R_{bs}(1-R_{bs})\sqrt{R_e} \langle p_{out} \rangle}{\sqrt{2}}. \qquad (6)$$

In a LOFI experiment, because the laser simultaneously fulfills the functions of the source and the detector, the saturation level is defined as the effective reflectivity corresponding to a maximum modulation of the laser output power (MC=1):

$$R_{Sat}(F_e) = \frac{1}{4} \times \frac{1}{G^2(F_e)} \times \frac{1}{(1-R_{bs})^2} \qquad (7)$$

which corresponds to the following value for the LOFI signal:

$$S_{Sat} = S_{LOFI}(R_{sat}, F_e) = \frac{R_{bs}\langle p_{out}\rangle}{\sqrt{2}}, \tag{8}$$

In this paper, whatever the experimental conditions (i.e. whatever the laser power and the amount of optical feedback), we supposed that the laser saturation level (MC=1) is always below the saturation level of the detection setup (photodiode and/or lock-in).

In a LOFI interferometer, a particularly interesting situation is the resonance case ($F_e = F_R$) where the LOFI signal gain (i.e. the autodyne gain) becomes:

$$G(F_R) = \frac{\gamma_c}{\gamma_1 \eta}. \tag{9}$$

Thus, for a LOFI setup, the most important parameter is the cavity damping rate to the population-inversion damping rate ratio ($\gamma_c/\gamma_1$). For a microchip laser, this ratio is typically of the order of $10^6$ [10] and the main advantage of the LOFI detection technique seems to come from this resonant amplification of the optical wave mixing [10, 15, 18]. For example, assuming a pumping rate of $\eta = 2$ and a 50/50 split ratio ($R_{bs} = 0.5$), a maximum output power modulation (MC=1) is obtained for a very weak feedback level corresponding to an effective reflectivity of $R_{Sat}(F_R) = 4 \times 10^{-12}$.

## *B. LOFI signal to noise ratio*

Without optical feedback ($\gamma_e = 0$), the set of Eqs. (1), allows us to study the laser quantum fluctuations induced by the Langevin noise terms ($F_N(t)$, $F_{E_c}(t)$ and $F_{\Phi_c}(t)$). For small fluctuations, Eqs. (1) can be linearized around its steady state which leads to the power density spectrum of the laser output power quantum fluctuations [18, 25]:

$$\text{PD}_{\text{Laser}}(F) = 2\langle p_{\text{out}} \rangle \frac{\gamma_c^2}{4\pi^2} \frac{(\Delta F_R^2 + F^2)}{(F_R^2 - F^2)^2 + \Delta F_R^2 F^2} = 2\langle p_{\text{out}}(t) \rangle G^2(F). \quad (10)$$

The LOFI noise power, detected by the photodiode (i.e. after the beam splitter reflection) and the lock-in amplifier, in a bandwidth $\Delta F$ around the feedback shift frequency $F_e$, is then given by:

$$N_{\text{Laser}}^2(F_e, \Delta F) = 2R_{\text{bs}} \int_{F_e - \Delta F/2}^{F_e + \Delta F/2} \text{PD}_{\text{Laser}}(F) dF. \quad (11)$$

If the detection bandwidth is narrower than the resonance width ($\Delta F \ll \Delta F_R$) the noise is then simply given by:

$$N_{\text{Laser}}(F_e, \Delta F) = 2\sqrt{R_{\text{bs}} \langle p_{\text{out}}(t) \rangle} G(F_e) \sqrt{\Delta F}. \quad (12)$$

At this point, one can notice that the resonant amplification gain $G(F_e)$ present in the LOFI signal [see Eq. (6)], is also present in the LOFI noise and, as a result, the signal to noise ratio (SNR) of the LOFI setup is frequency independent:

$$\frac{S_{\text{LOFI}}(R_e, F_e)}{N_{\text{Laser}}(F_e, \Delta F)} = \frac{\sqrt{R_{\text{bs}} \langle p_{\text{out}} \rangle}}{\sqrt{2\Delta F}} \times \sqrt{R_e}(1 - R_{\text{bs}}), \quad (13)$$

and, as mentioned above, the relaxation frequency seems to be of no particular importance.

For example, with a laser output power $P_{out} = 10\,\text{mW}$ ($\langle p_{out} \rangle = 5.36 \times 10^{16}$ photons/s at $\lambda = 1064\,\text{nm}$), a detection bandwidth $\Delta F = 1\,\text{kHz}$ and a beam splitter reflectivity $R_{bs} = 0.5$, whatever the shift frequency may be (resonant or non-resonant), the minimum measurable reflectivity value (resulting in SNR=1) is equal to:

$$R_{Laser}(\Delta F) = \frac{2\Delta F}{R_{bs}\langle p_{out}\rangle} \times \frac{1}{(1-R_{bs})^2} = 3 \times 10^{-13} \,. \tag{14}$$

Physically, Eq. (14) means that, during the integration time ($T = \frac{1}{2\Delta F}$), only $1/R_{bs}$ reflected photons are detected:

$$\frac{R_{Laser}(1-R_{bs})^2\langle p_{out}\rangle}{2\Delta F} = \frac{1}{R_{bs}} \,. \tag{15}$$

Therefore, in the LOFI setup, the beam splitter reflectivity ($R_{bs}$) appearing in Eqs. (6) and (11), and finally in the right hand term of Eq. (15), can be interpreted as the quantum efficiency of the LOFI detection (indirect detection by a photodiode).

For a laser output power $P_{out} = 8.8\,\text{mW}$ (i.e. $\langle p_{out} \rangle = 4.7 \times 10^{16}$ photons/s at $\lambda = 1064\,\text{nm}$), Figs. 2(a) and 2(b) show respectively the power spectra of the LOFI signal $S^2_{LOFI}(R_e, F_e)$ and of the LOFI noise $N^2_{Laser}(F_e, \Delta F)$. The power spectra are normalized (i.e. rescaled) by dividing them by the quantum shot noise level ($\langle p_{out} \rangle 2\Delta F$). As a result, the 0dB level corresponds to the shot noise level. As already mentioned, the LOFI signal and the LOFI noise exhibit the same resonance and as a result, the signal to noise ratio of the LOFI setup is frequency independent and is given by: $\frac{S_{LOFI}^2 \left( R_e = 2 \times 10^{-11}, \Omega_e \right)}{N_{Laser}^2 \left( F_e, 600\,Hz \right)} = 100$ (i.e. 20 dB).

Now, by taking into account the detection noise which is assumed to be a white noise [see Fig. 2(d)], the signal to noise ratio of the LOFI setup is shot noise limited only within a frequency range close to the laser relaxation frequency ( $F_- \leq F_e \leq F_+$ ) where the laser quantum noise [Fig. 2(c)] is higher than the detection noise [Fig. 2(d)]. At this point, one can understand that the main advantages of the resonant amplification of the LOFI signal and of the LOFI noise are to obtain, in a very simple way, with a conventional photodiode and with no complex alignment, a shot noise limited setup for interferometric measurements.

## C. Optimization of the LOFI measurement dynamic range

To get a LOFI signal which is shot noise limited and also not saturated, the following inequalities need to be satisfied:

$$N_{Detector}^2 (\Delta F) \leq N_{Laser}^2 (F_e, \Delta F) \leq S_{LOFI}^2 (R_e, F_e) \leq S_{Sat}^2, \qquad (16)$$

where we assume a white detection noise [Fig. 2(d)], with the following level (in photon units):

$$N_{Detector}(\Delta F) = \left[ \frac{\left( 6 \times 10^{-9}\,W/\sqrt{Hz} \right)}{hc/\lambda} \right] \sqrt{2 \Delta F}. \qquad (17)$$

By using Eq. (6) to replace the laser output power modulation by the effective reflectivity, Eq. (16) can be rewritten:

$$R_{Detector}(F_e, \Delta F) \leq R_{Laser}(\Delta F) \leq R_e \leq R_{Sat}(F_e) \qquad (18)$$

where $R_{Sat}(F_e)$ is given by Eq. (7), $R_{Laser}(\Delta F)$ is given by Eq. (14) and where:

$$R_{Detector}(F_e, \Delta F) = \frac{1}{2} \times \frac{1}{G^2(F_e)} \times \frac{1}{(1-R_{bs})^2 R_{bs}^2} \times \frac{N^2_{detector}(\Delta F)}{\langle p_{out} \rangle^2}. \tag{19}$$

From the point of view of the detection, the best LOFI configuration in terms of sensitivity and dynamic range, is obtained when the shift frequency $F_e$ is equal to $F_+$ (or $F_-$), i.e. one of the values where, the laser noise intercepts the white detection noise on Fig. 2. Indeed for $F_e = F_+$ (or $F_e = F_-$), and with the experimental conditions used to obtain the power spectra shown on Fig. 2, one obtains:

$$R_{Laser}(600\,Hz) = 2 \times 10^{-13} \leq R_e \leq R_{Sat}(\Omega_+) = 4.5 \times 10^{-5}. \tag{20}$$

In these conditions, the LOFI setup is therefore shot noise limited [see Eq. (15)], and has a dynamic range of $84\,dB$, given by the ratio $R_{Sat}/R_{Laser}$.

For $F_- < F_e = F_R < F_+$ : one obtains:

$$R_{Laser}(600\,Hz) = 2 \times 10^{-13} \leq R_e \leq R_{Sat}(F_R) = 4 \times 10^{-8}. \tag{21}$$

The LOFI setup is therefore still shot noise limited, but it has a lower dynamic range ($53\,dB$), induced by the resonant amplification of the LOFI signal and of the LOFI noise which are both closer to the saturation level [see Fig. 2].

Finally, for $F_e = 1.5 \times F_R > F_+$ the LOFI detection is now limited by the detection noise:

$$R_{detector}(1.5 \times F_R, 600\,Hz) = 3 \times 10^{-13} \leq R_e \leq R_{sat}(1.5 \times F_R) = 7 \times 10^{-5}. \tag{22}$$

In these conditions, the LOFI setup is no more shot noise limited, but it still has the highest dynamic range ($83\,\text{dB}$).

Otherwise, by working at the resonance frequency ($F_e = F_R$), the smallest reflectivity measurable with the LOFI setup can be determined by writing:

$$N_{\text{Detector}}^2(\Delta F) = N_{\text{Laser}}^2(F_R, \Delta F) \tag{23}$$

which is equivalent to:

$$R_{\text{Detector}}(F_R, \Delta F) = R_{\text{Laser}}(\Delta F). \tag{24}$$

Using Eq. (23) [or Eq. (24)], the optimum value for the LOFI gain can be established:

$$G_{\text{opt}}(F_R) = \left(\frac{\gamma_c}{\gamma_1 \eta}\right)_{\text{opt}} = \frac{N_{\text{Detector}}(\Delta F)}{2\sqrt{R_{\text{bs}}\langle p_{\text{out}}\rangle \Delta F}} = \frac{1}{\sqrt{2R_{\text{bs}}}} \frac{\left[\frac{(6\times 10^{-9}\,\text{W}/\sqrt{\text{Hz}})}{hc/\lambda}\right]}{\sqrt{\langle p_{\text{out}}\rangle}}. \tag{25}$$

The LOFI gain of the optimum laser is thus simply given by the ratio between the power density spectra of the detection noise level and the shot noise level.

For the same output power, the same laser relaxation frequency and the same experimental detection conditions, Fig. 2 and Fig. 3 allow a comparison of the power spectra of the LOFI signal, the LOFI noise and the detection noise obtained for two different values of the LOFI gain. In Fig. 3, the LOFI gain is optimum ($\left(\frac{\gamma_c}{\gamma_1 \eta}\right)_{\text{opt}} = 146$) and the best shift frequency is the relaxation frequency where the LOFI detection is shot noise limited and has also the highest dynamic range:

$$R_{Laser}(600\,\text{Hz}) = 2 \times 10^{-13} \leq R_e \leq R_{sat}(F_R) = 4.5 \times 10^{-5} \qquad (26)$$

while in Fig. 2 the LOFI gain is higher than the optimum value $\left(\dfrac{\gamma_c}{\gamma_1 \eta}\right) = 5 \times 10^3 > \left(\dfrac{\gamma_c}{\gamma_1 \eta}\right)_{opt}$ and, as already mentioned, the best shift frequency (in terms of sensitivity and dynamic range for the LOFI sensor) is not the laser relaxation oscillation frequency, but $F_+$ (or $F_-$). At this point one can verify that for this frequency the gain is also equal to the optimum gain: $G(F_+) = \left(\dfrac{\gamma_c}{\gamma_1 \eta}\right)_{opt}$.

To drive the point home, let's notice that the use of a microchip laser, with a very high value of the LOFI gain parameter $\left(\dfrac{\gamma_c}{\gamma_1 \eta}\right) = 5 \times 10^5 \gg \left(\dfrac{\gamma_c}{\gamma_1 \eta}\right)_{opt}$ as used in Refs. [10] or [15] is not so interesting. Indeed, a high gain is interesting from the signal point of view, which is resonantly amplified [see Eq. (6)] but also from the detection point of view which is shot noise limited [see Eq. (15)], because the resonant quantum noise of the laser is above the detection noise. Unfortunately, a high gain is uninteresting for the dynamic range of the detection which is for example very low (13 dB) when working at the resonance frequency, due to the fact that the laser noise level is very close to the laser saturation level:

$$R_{Laser}(600\,\text{Hz}) = 2 \times 10^{-13} \leq R_e \leq R_{sat}(F_R) = 4 \times 10^{-12}. \qquad (27)$$

With this kind of laser having a very high LOFI gain, to obtain the maximum value of the dynamic range (83 dB), the best shift frequency is very far from the resonance ($F_+ \approx 7 \times F_R$) and the high resonant amplification effect seems to be unimportant [26].

To conclude this section, we have numerically compared the dynamical behavior of two lasers having the same output power and the same relaxation frequency, but having different values of the LOFI gain ($G(F_R) = \gamma_c / \gamma_1 \eta$). Figure 4 shows 1D scans extracted from the measured output power modulation obtained from the numerical integration of the set of Eqs. (1) by a Runge-Kutta method. The target under investigation is a reflectivity stair composed of four steps. The first step ($R_e = 0$) allows to the observe the effect of the Langevin noise terms (i.e. the effect of the Laser quantum noise), while the other steps ($R_e \neq 0$) allow observation of the effects of increasing effective reflectivities combined with the laser noise.

For the laser, having the lower value of the LOFI gain ($G(F_R) = 1 \times 10^4$), the numerical results shown on Fig. 4(c) are in good agreement with the analytical predictions. Indeed, when the effective reflectivity is multiplied by a factor 100 (step n°2 to step n°4), the modulation contrast increases by a factor 10, while when the effective reflectivity is multiplied by a factor 4 (step n°3 to step n°4) the modulation contrast increases by a factor 2. Moreover, in Table 1, the modulation contrast (MC) and the signal to noise ratio SNR, numerically determined using the results of Fig. 4(c) and analytically calculated [from Eqs. (4) and (13)], are very close. One can notice that the results are in good agreement because the experimental conditions are below the saturation level ($R_{e,4} = 10^{-10} < R_{sat}(F_R) = 2.5 \times 10^{-9}$). With the same laser ($G(F_R) = 1 \times 10^4$), by increasing the detection bandwidth [Figs. 4(d)], one can observe that the values of the modulation contrast are always the same but are now accompanied by an increase of the noise level. As a result, the second step ($R_{e,2} = 10^{-12}$) of the reflectivity stair, is now very close to the noise level. For this bandwidth ($\Delta F = 66.6\,\text{kHz}$) the modulation contrast (MC) and the signal to noise ratio (SNR) calculated analytically and determined numerically [from Fig. 4(d)] are again in good agreement (see Tab. 1). Finally, in Table 1, the comparison of the results obtained in the low speed configuration ($\Delta F = 666\,\text{Hz}$) with the results obtained in the high speed configuration ($\Delta F = 66.6\,\text{kHz}$), confirms that the SNR decreases by a factor 10 when the detection bandwidth increases by a factor 100.

Now, by increasing the value of the LOFI gain ($G(F_R) = 5 \times 10^5$), one can observe [Fig. 4(a) or 4(b)], that the modulation contrast is higher. This result is in agreement with the theoretical prediction of Eq. (4), where the modulation contrast increases with the value of the LOFI gain. By comparing Figs. 4(a) and 4(b) with respectively Figs. 4(c) and 4(d), one can also observe that the signal to noise ratio is lower with the laser having the higher LOFI gain, independently of the detection bandwidth ($\Delta F = 666\,\text{Hz}$ or $\Delta F = 66.6\,\text{kHz}$). By increasing the value of the LOFI gain, the degradation of the signal to noise ratio comes from both the resonant amplification of the laser noise [see Eq. (12)] and the saturation of the LOFI signal. The saturation effect is clearly visible if we compare, for example, the relative height of the second and fourth steps on Fig. 4(a). Indeed, for these two steps, the effective reflectivity is multiplied by a factor 100 while the modulation contrast increases only by a factor 3. The saturation effect is also visible in Table 2 where for both SNR and MC, the numerical results [obtained from Fig. 4(a) and 4(b)], are always lower than the analytical ones [obtained from Eqs. (4) and (13)]. At this point one can notice that the saturation effect observed numerically, can not be obtained analytically due to the fact that Eq (4), and finally Eq. (13), are obtained after a linearization of the set of Eqs. (1), i.e. far way from the saturation conditions [18]. Moreover, one can also notice that the condition ($\Delta F \ll \Delta F_R$), mentioned for the derivation of Eq. (13), is not satisfied for the numerical simulation shown on Fig. 4(b) ($\Delta F = 66.6\,\text{kHz}$ and $\Delta F_R = 3\,\text{kHz}$).

Finally the numerical images shown on Figs. 4, and the comparison of the numerical and of the analytical values of MC and SNR shown in Tables 1 and 2, allow to conclude that the higher signal (i.e. the higher modulation contrast) is obtained for the laser with higher value of the LOFI gain, but that the best images (i.e. the highest SNR) is obtained when using the laser with the lower value of the LOFI gain.

## 4. MICHELSON INTERFEROMETER AGAINST LOFI INTERFEROMETER

The main objective of this section is to compare the signal to noise ratio of a Michelson interferometer (i.e. a heterodyne setup) with the one of a LOFI interferometer (i.e. an autodyne setup).

In the autodyne setup [Fig. 1(b)], the power (number of photons) of the reference wave which goes to the detector is given by:

$$p_{ref} = R_{bs}(1 - R_{bs})\langle p_{out} \rangle \tag{28}$$

while the power of the signal wave obeys to:

$$p_{sig} = R_{bs}(1 - R_{bs})R_e\langle p_{out} \rangle \tag{29}$$

and finally, the demodulation of the heterodyne output power modulation by the means of a lock-in amplifier allows us to obtain the heterodyne signal:

$$S_{Hetero}(R_e) = \frac{2\sqrt{p_{ref} p_{sig}}}{\sqrt{2}} = \frac{2R_{bs}(1-R_{bs})\sqrt{R_e}\langle p_{out}\rangle}{\sqrt{2}}. \tag{30}$$

which is supposed to be always below the saturation level of the detection setup (photodiode and lock-in).

The comparison of the heterodyne signal given by Eq. (30) with the autodyne signal given by Eq. (6), shows that the LOFI signal is amplified by the gain $G(F_e)$, with a maximum value given by $G(F_R)$ when working at the resonance frequency. Assuming that the detection of the heterodyne setup is limited by the detection noise [Eq. (17)], we can define the signal to noise ratio:

$$\frac{S_{Hetero}(R_e)}{N_{Detector}(\Delta F)} = \frac{R_{bs}(1-R_{bs})\sqrt{R_e}\langle p_{out}\rangle}{\left(\frac{6\times 10^{-9}\, W/\sqrt{Hz}}{hc/\lambda}\right)\sqrt{\Delta F}}. \tag{31}$$

With the experimental conditions ($P_{out} = 8.8\,mW$ and $\Delta F = 600\,Hz$) corresponding to Figs. 2 and 3, and using Eq. (31), one obtains a minimum effective reflectivity (SNR=1) of $2.23\times 10^{-9}$ which gives a heterodyne dynamic range of 87 dB a little bit higher than the best LOFI dynamic range of the LOFI setup (83 dB).

The comparison of the signal to noise ratio of the autodyne setup [Eq. (13)] and of the heterodyne setup [Eq. (31)] allows us to determine the real gain of the LOFI interferometer compared to the Michelson interferometer:

$$\frac{S_{LOFI}(R_e, F_e)}{N_{Laser}(F_e, \Delta F)} \bigg/ \frac{S_{Hetero}(R_e)}{N_{Detector}(\Delta F)} = \frac{1}{\sqrt{2R_{bs}}} \frac{\left[\frac{(6 \times 10^{-9}\, W/\sqrt{Hz})}{hc/\lambda}\right]}{\sqrt{\langle p_{out} \rangle}} = G_{opt}(\Omega_R). \tag{32}$$

One can notice that, below the LOFI saturation level (as long as $R_e \leq R_{Sat}(F_e)$), whatever the laser may be (as long as the following condition is satisfied: $G(F_R) \geq G_{opt}(F_R)$) and whatever the shift frequency may be (as long as $F_- \leq F_e \leq F_+$), the real gain of the LOFI setup is therefore given by the optimum value $G_{opt}(F_R)$ of the LOFI gain.

For example, on Fig. 2, and for $F_- \leq F_e \leq F_+$, where the laser quantum noise is above the detection noise, we have for the LOFI detection:

$$\frac{S_{LOFI}^2(R_e = 2 \times 10^{-11}, F_e)}{N_{Laser}^2(F_e, \Delta F = 600\,Hz)} = +20\,dB, \tag{33}$$

which means that the LOFI signal is ten times more important than the laser quantum noise. Respectively, for the heterodyne detection using the Michelson interferometer, we have:

$$\frac{S_{Hetero}^2(R_e = 2 \times 10^{-11})}{N_{Detector}^2(\Delta F = 600\,Hz)} = -23.3\,dB, \tag{34}$$

which means that the heterodyne signal is below the detection noise and therefore that the heterodyne detection of the target having an effective reflectivity of $R_e = 2 \times 10^{-11}$ is impossible using this detection bandwidth. Finally, in agreement with the analytical prediction, one can determine from Eqs. (33) and (34) that the real advantage of the LOFI interferometer can be quantified by the optimum value of the LOFI gain:

$$\frac{S_{LOFI}(R_e, F_e)}{N_{Laser}(F_e, \Delta F)} \bigg/ \frac{S_{Hetero}(R_e)}{N_{Detector}(\Delta F)} = 21.65\,\text{dB} = 146 = G_{opt}(\Omega_R) \qquad (35)$$

and this despite the fact that the LOFI experiment has been conducted with a laser having a very high LOFI gain compared to the optimum value: $G(F_R) = 5 \times 10^3 \gg G_{opt}(F_R)$.

Using Eq. (32), one can compare the LOFI interferometer (autodyne setup) and the heterodyne interferometer in a more general way. Indeed, for a detection noise level of $6 \times 10^{-9}\,\text{W}/\sqrt{\text{Hz}}$, we obtain the same value of the signal to noise ratio (i.e. $G_{opt}(F_R) = 1$) for the two kinds of interferometers, if the laser output power is equal to 193 mW (i.e. $\langle p_{out} \rangle = 1 \times 10^{18}$ photons/s when working at $\lambda = 1064\,\text{nm}$). In these conditions, both the heterodyne and the autodyne setup are shot noise limited [Eq. (16)]. Likewise, for a laser output power $P_{out} = 8.8\,\text{mW}$, one obtains $G_{opt}(F_R) = 1$, for a detection noise level as low as: $4 \times 10^{-11}\,\text{W}/\sqrt{\text{Hz}}$. Finally, we can conclude that compare to a heterodyne interferometer, the LOFI detection setup (autodyne interferometer) is competitive ($G_{opt}(F_R) > 1$) when working at a low power level, with a conventional noisy detection.

## 5. CONCLUSION

For a LOFI interferometer, involving a laser having relaxation oscillations, we have recalled that both the signal and the laser quantum noise can exhibit a strong amplification when the waves beating is realized at the resonance frequency (i.e. the laser relaxation frequency). The main advantage of the resonant amplification is to allow the laser quantum noise to be above the detection noise in a frequency range close the laser relaxation frequency. Under these conditions, the signal to noise ratio of a LOFI setup is frequency independent and most importantly shot noise limited.

To maximize the dynamic range of the LOFI setup, we have determined that the best value of the shift frequency is the frequency at which the laser quantum noise equals the detection noise level. Equivalently, when operating at the relaxation frequency, we have determined that the optimum value of the LOFI gain is simply given by the ratio between the power density levels of the detection noise and of the shot noise. In these conditions the dynamic range of an heterodyne interferometer and of an autodyne interferometer are roughly equivalent. We have also numerically demonstrated that a too high LOFI gain (compare to the optimum value) induces a decrease of the image quality (signal to noise and also contrast) by saturation effects.

For given laser output power, target under investigation and detection noise level, we have compare the signal to noise ratio of a LOFI setup (autodyne interferometer) with a conventional Michelson interferometer (heterodyne interferometer). Irrespective of the LOFI gain (defined by the ratio between the cavity damping rate and the population-inversion damping rate of the laser), we have demonstrated that the real performance gain of an autodyne setup compared to a heterodyne setup is simply given by the optimum value of the LOFI already mentioned. From this study we have conclude that compare to a heterodyne interferometer, the LOFI detection setup (autodyne interferometer) is competitive (sensitivity, dynamic range) when the optimum value og the LOFI gain is greater than unity i.e. when working at a low laser power level, and/or with a conventional noisy detection.

Finally, one can also recall that, compared to the Michelson interferometer, the LOFI setup is always self-aligned and therefore is much more robust because it doesn't need any complex alignment.

# REFERENCES

1. T. Yoshizawa, editor, *Handbook of optical metrology: Principles and Applications* (CRC Press, 2009).

2. T. Sawatari, "Optical heterodyne scanning microscope," Appl. Opt. **12**, 2768-2772 (1973).

3. J.A. Izatt, M. R. Hee, G. M. Owen, E. A. Swanson, and J. G. Fujimoto, "Optical coherence microscopy in scattering media," Opt. Lett. **19**, 590-592 (1994).

4. A.D. Aguirre, J. Sawinski, S. W. Huang, C. Zhou, W. Denk, and J. G. Fujimoto, "High speed optical coherence microscopy with autofocus adjustment and a miniaturized endoscopic imaging probe," Opt. Express **18**, 4222-4239 (2010).

5. C. Zhou, T. H. Tsai, D. Adler, H. C. Lee, D. W. Cohen, A. Mondelblatt, Y. Wang, J. L. Connolly, and J. G. Fujimoto, "Photothermal optical coherence tomography in ex vivo human breast tissues using gold nanoshells," Opt. Lett. 35 700-702 (2010).

6. V. J. Srinivasan, S. Sakadzic, I. Gorczynska, S. Ruvinskaya, W. Wu, J.G. Fujimoto, and D.A. Boas, "Quantitative cerebral blood flow with Optical Coherence Tomography," Opt. Express, **18**, 2477 2494 (2010).

7. M. Kempe, W. Rudolph, and E. Welsh, "Comparative study of confocal heterodyne microscopy for imaging through scattering media," J. Opt. Soc. Am. A **13**, 46-52 (1996).

8. Y. Niwa, K. Arai, A. Ueda, M. Sakagami, N. Gouda, Y. Kobayashi, Y. Yamada, and T. Yano, "Long-term stabilization of a heterodyne metrology interferometer down to noise level of 20 pm over an hour," Appl. Opt. **48**, 6105-6110 (2009).

9. I. Hahn, M. Xeilert, X. Wang, and R. Goullioud, "A heterodyne interferometer for angle metrology," Rev. Sci. Instrum. **81**, 045103 (2010).

10. K. Otsuka, "Highly sensitive measurement of Doppler-shift with a microchip solid-state laser," Jpn. J. Appl. Phys. **31,** L1546–L1548 (1992).

11. S. Okamoto, H. Takeda, and F. Kannari, "Ultrahighly sensitive laser-Doppler velocity meter with a diode-pumped Nd:YVO4 microchip laser," Rev. Sci. Instrum. **66,** 3116–3120 (1995).

12. K. Otsuka, K. Abe, J.Y. Ko, and T.S. Lim, "Real-time nanometer vibration measurement with self-mixing microchip solid-state laser," Opt. Lett. **27**, 1339-1341 (2002).

13. H. Gilles, S. Girard, M. Laroche, and A. Belarouci, "Near-field amplitude and phase measurements using heterodyne optical feedback on solid-state lasers," Opt. Lett. **33**, 1-3 (2008).

14. S. Blaize, B. Bérenguier, I. Stéfanon, A. Bruyant, G. Lerondel, P. Royer, O. Hugon, O. Jacquin, and E. Lacot, Opt. Express **16**, 11718-11726 (2008).

15. E. Lacot, R. Day, and F. Stoeckel, "Laser optical feedback tomography," Opt. Lett. **24,** 744–746 (1999).

16. A. Witomski, E. Lacot, O. Hugon, and O. Jacquin, "Synthetic aperture laser optical feedback imaging using galvanometric scanning," Opt. Lett. 31, 3031-3033 (2006).

17. O. Jacquin, S. Heidmann, E. lacot, and O.Hugon, "Self aligned setup for laser optical feedback imaging insensitive to parasitic optical feedback" Appl. Opt. **48**, 64-68 (2009)

18. E. Lacot, R. Day, and F. Stoeckel, "Coherent laser detection by frequency-shifted optical feedback," Phys. Rev. A **64,** 043815 (2001).

19. E. Lacot, and O. Hugon, "Phase-sensitive laser detection by frequency-shifted optical feedback," Phys. Rev. A **70,** 053824 (2004).

20. J.J. Zaykowski and A. Mooradian, "Single –frequency microchip Nd lasers," Opt. Lett. 14, 24-26 (1989)

21. O. Hugon, I.A. Paun, C. Ricard, B. van der Sanden, E. Lacot, O. Jacquin, A. Witomski, "Cell imaging by coherent backscattering microscopy using frequency shifted optical feedback in a microchip laser," Ultramicroscopy **108**,523-528 (2008).

22. V. Muzet, E.Lacot, O.Hugon, Y. Gaillard, "Experimental comparison of shearography and laser optical feedback imaging for crack detection in concrete structures," Proc. SPIE 5856, 793-799 (2005).

23. M. Sargent III, M. O. Scully, and W. E. Lamb, *Laser Physics* (Addison-Wesley, Reading, MA, 1974).

24. K. Petermann, *Laser Diode Modulation and Noise* (Kluwer Academic, Dordrecht, 1991).

25. M.I. Kolobov, L. Davidovich, E. Giacobino, and C. Fabre, "Role of pumping statistics and dynamics of atomic polarization in quantum fluctuations of laser sources," Phys. Rev. A **47**, 1431-1446 (1993).

26. In Ref. (19), we have demonstrated that, far way from resonance, the nonlinear phase noise of a LOFI setup is less important. The working condition ($F_+ \approx 7 \times F_R$) could be interesting to obtain a shot noise limited setup for phase measurements with a very high precision level.

**FIGURE CAPTIONS**

Fig. 1. Schematic diagrams of the autodyne interferometer setup (a) and of the heterodyne interferometer setup (b) for scanning microscopy. $L_1$, $L_2$ and $L_3$: Lenses, OI: Optical Isolator BS: Beam Splitter with a power reflectivity $R_{bs}$, GS: Galvanometric Scanner, RM: Reference Mirror with a unitary power reflectivity $R_{rm}=1$, FS Frequency Shifter with a round trip frequency-shift $F_e$, PD: Photodiode with a white noise spectrum. The lock-in amplifier is characterized by a bandwidth $\Delta F$ around the reference frequency $F_e$. The laser output power is characterized by $p_{out}$ (photons/s), the target is characterized by its effective reflectivity $R_e \ll 1$.

Fig. 2. Normalized power spectra of the laser output power versus the normalized shift frequency. The Power spectra are normalized to the quantum shot noise with $\Delta F = 600\,\text{Hz}$ and $p_{out} = 4.7 \times 10^{16}\,\text{photons/s}$ ($P_{out} = 8.8\,\text{mW}$ at $\lambda = 1064\,\text{nm}$): a) saturation level, b) autodyne signal ($R_e = 2 \times 10^{-11}$ ; $R_{bs} = 1/2$) c) autodyne noise (laser quantum noise), d) heterodyne noise (detection noise), e) heterodyne signal ($R_e = 2 \times 10^{-11}$ ; $R_{bs} = 1/2$). Laser dynamical parameters corresponding to a conventional Nd$^{3+}$:YAG microchip laser: $\gamma_c / \gamma_1 \eta = 5 \times 10^3$, $\eta = 2$, $\gamma_c = 5 \times 10^8\,\text{s}^{-1}$, $\gamma_1 = 5 \times 10^4\,\text{s}^{-1}$, $F_R = 796\,\text{kHz}$.

Fig. 3. Normalized power spectra of the laser output power versus the normalized shift frequency. The Power spectra are normalized to the quantum shot noise with $\Delta F = 600\,\text{Hz}$ and $p_{out} = 4.7 \times 10^{16}\,\text{photons/s}$ ($P_{out} = 8.8\,\text{mW}$ at $\lambda = 1064\,\text{nm}$): a) saturation level, b) autodyne signal ($R_e = 2 \times 10^{-11}$ ; $R_{bs} = 1/2$) c) autodyne noise (laser quantum noise), d) heterodyne noise (detection noise), e) heterodyne signal ($R_e = 2 \times 10^{-11}$ ; $R_{bs} = 1/2$). Laser dynamical parameters corresponding to the optimum values: $\gamma_c / \gamma_1 \eta = 146$, $\eta = 2$, $\gamma_c = 8.6 \times 10^7\,\text{s}^{-1}$, $\gamma_1 = 2.9 \times 10^5\,\text{s}^{-1}$, $F_R = 796\,\text{kHz}$.

Fig. 4. Numerical 1D scans obtained from the measured laser output power modulation contrast of an autodyne interferometer, when the laser beam is scanned on reflectivity stairs composed of 4 steps. Experimental conditions: Laser output power, $P_{out} = 117\,\text{mW}$ ($p_{out} = 6.25 \times 10^{17}\,\text{photons/s}$ at $\lambda = 1064\,\text{nm}$); Laser relaxation frequency, $F_R = 1.6\,\text{MHz}$; Autodyne modulation frequency, $F_e = F_R$. Step 1: (pixels 1 to 10), $R_{e,1} = 0$; Step 2: (pixels 11 to 20), $R_{e,2} = \dfrac{10^{-10}}{100}$ Step 3: (pixels 21 to 30), $R_{e,3} = \dfrac{10^{-10}}{4}$; Steps 4: (Pixels 31 to 40), $R_{e,4} = 10^{-10}$.

Top row: $G(F_R) = 5 \times 10^5$; Bottom row: $G(F_R) = 1 \times 10^4$; Left column: $\Delta F = 1/(2T) = 666\,\text{Hz}$; Right column): $\Delta F = 1/(2T) = 66.6\,\text{kHz}$.

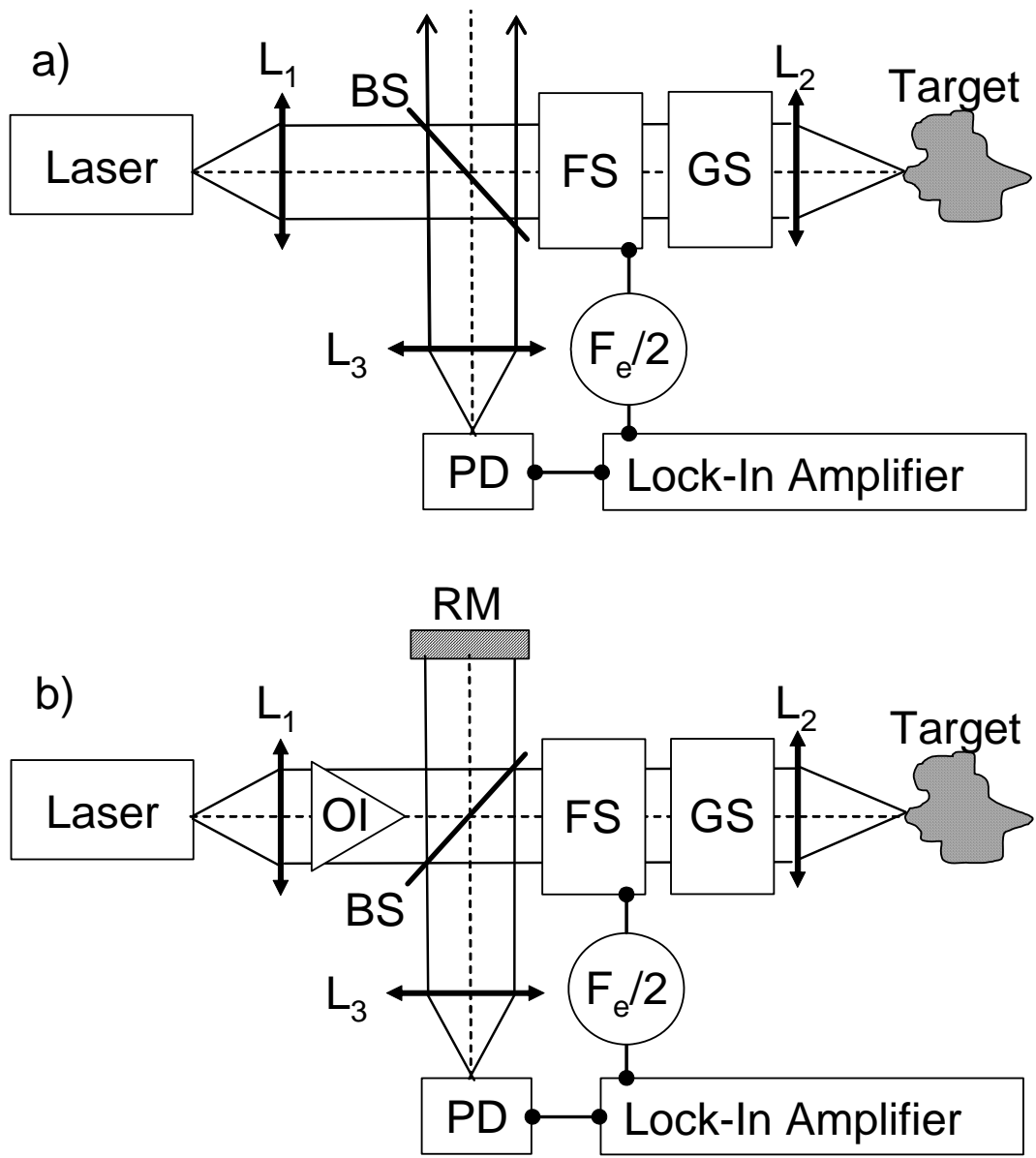

Fig. 1. Schematic diagrams of the autodyne interferometer setup (a) and of the heterodyne interferometer setup (b) for scanning microscopy. $L_1$, $L_2$ and $L_3$: Lenses, OI: Optical Isolator BS: Beam Splitter with a power reflectivity $R_{bs}$, GS: Galvanometric Scanner, RM: Reference Mirror with a unitary power reflectivity $R_{rm}=1$, FS Frequency Shifter with a round trip frequency-shift $F_e$, PD: Photodiode with a white noise spectrum. The lock-in amplifier is characterized by a bandwidth $\Delta F$ around the reference frequency $F_e$. The laser output power is characterized by $p_{out}$ (photons/s), the target is characterized by its effective reflectivity $R_e \ll 1$.

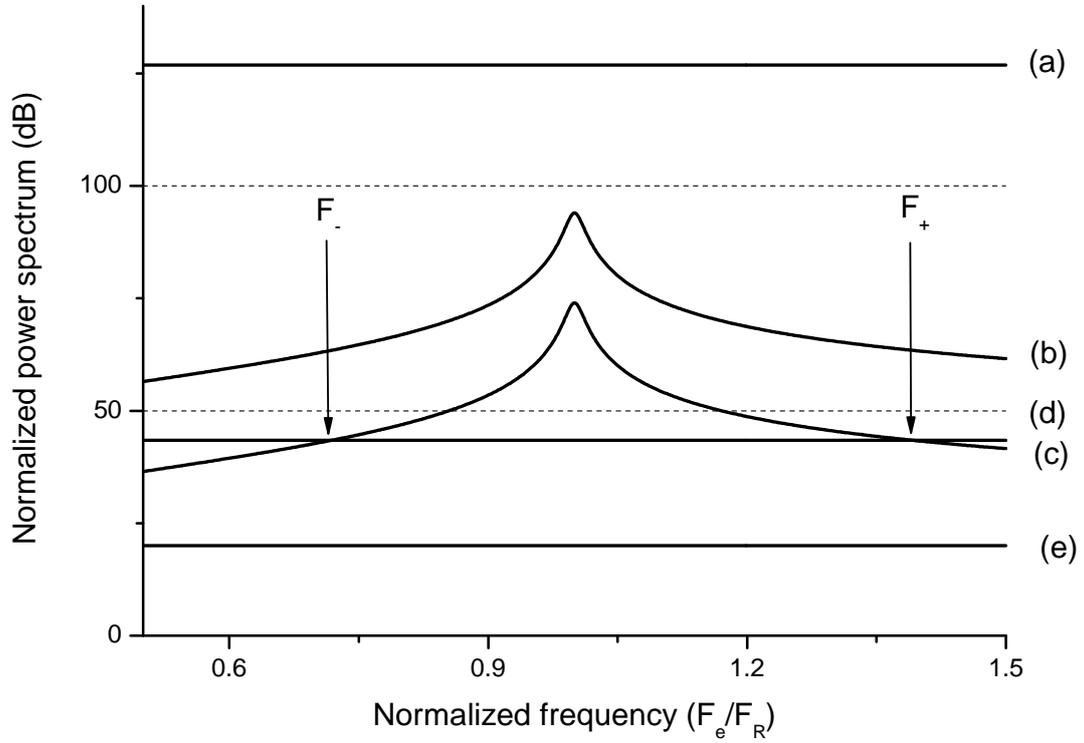

Fig. 2. Normalized power spectra of the laser output power versus the normalized shift frequency. The Power spectra are normalized to the quantum shot noise with $\Delta F = 600\,\text{Hz}$ and $p_{out} = 4.7 \times 10^{16}\,\text{photons/s}$ ($P_{out} = 8.8\,\text{mW}$ at $\lambda = 1064\,\text{nm}$): a) saturation level, b) autodyne signal ($R_e = 2 \times 10^{-11}$ ; $R_{bs} = 1/2$) c) autodyne noise (laser quantum noise), d) heterodyne noise (detection noise), e) heterodyne signal ($R_e = 2 \times 10^{-11}$ ; $R_{bs} = 1/2$). Laser dynamical parameters corresponding to a conventional $Nd^{3+}$:YAG microchip laser: $\gamma_c/\gamma_1\eta = 5 \times 10^3$, $\eta = 2$, $\gamma_c = 5 \times 10^8\,\text{s}^{-1}$, $\gamma_1 = 5 \times 10^4\,\text{s}^{-1}$, $F_R = 796\,\text{kHz}$.

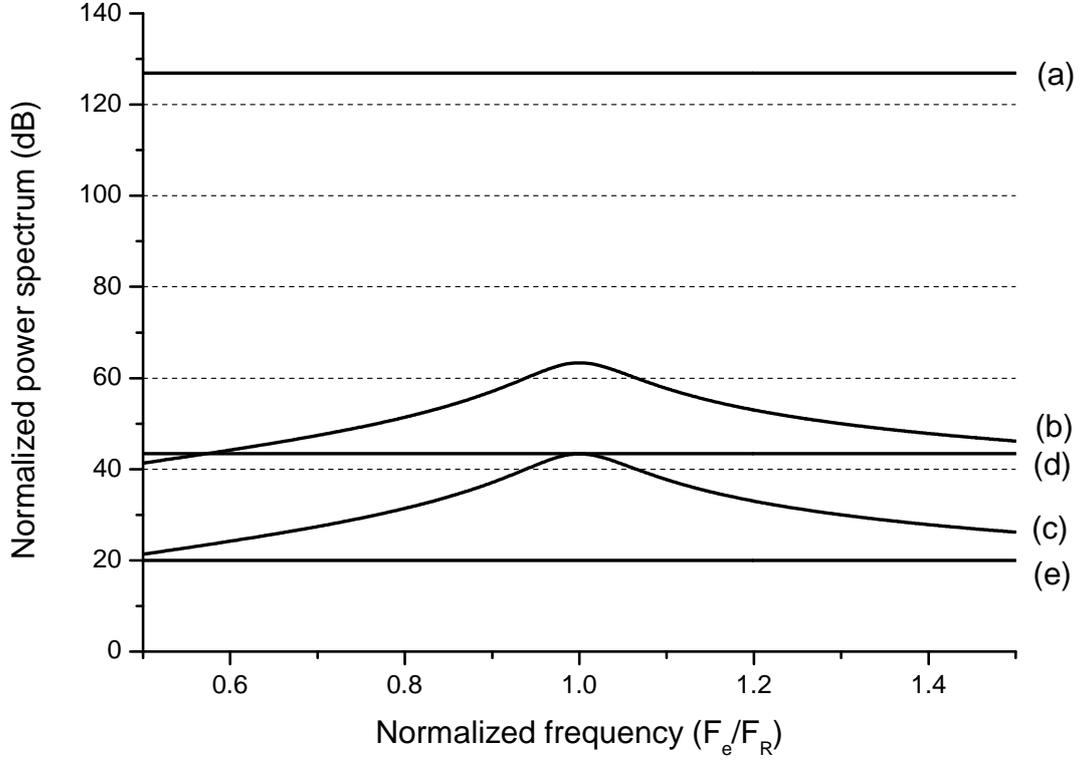

Fig. 3. Normalized power spectra of the laser output power versus the normalized shift frequency. The Power spectra are normalized to the quantum shot noise with $\Delta F = 600\,\text{Hz}$ and $p_{out} = 4.7 \times 10^{16}\,\text{photons/s}$ ($P_{out} = 8.8\,\text{mW}$ at $\lambda = 1064\,\text{nm}$): a) saturation level, b) autodyne signal ($R_e = 2 \times 10^{-11}$ ; $R_{bs} = 1/2$) c) autodyne noise (laser quantum noise), d) heterodyne noise (detection noise), e) heterodyne signal ($R_e = 2 \times 10^{-11}$ ; $R_{bs} = 1/2$). Laser dynamical parameters corresponding to the optimum values: $\gamma_c / \gamma_1 \eta = 146$, $\eta = 2$, $\gamma_c = 8.6 \times 10^7\,\text{s}^{-1}$, $\gamma_1 = 2.9 \times 10^5\,\text{s}^{-1}$, $F_R = 796\,\text{kHz}$.

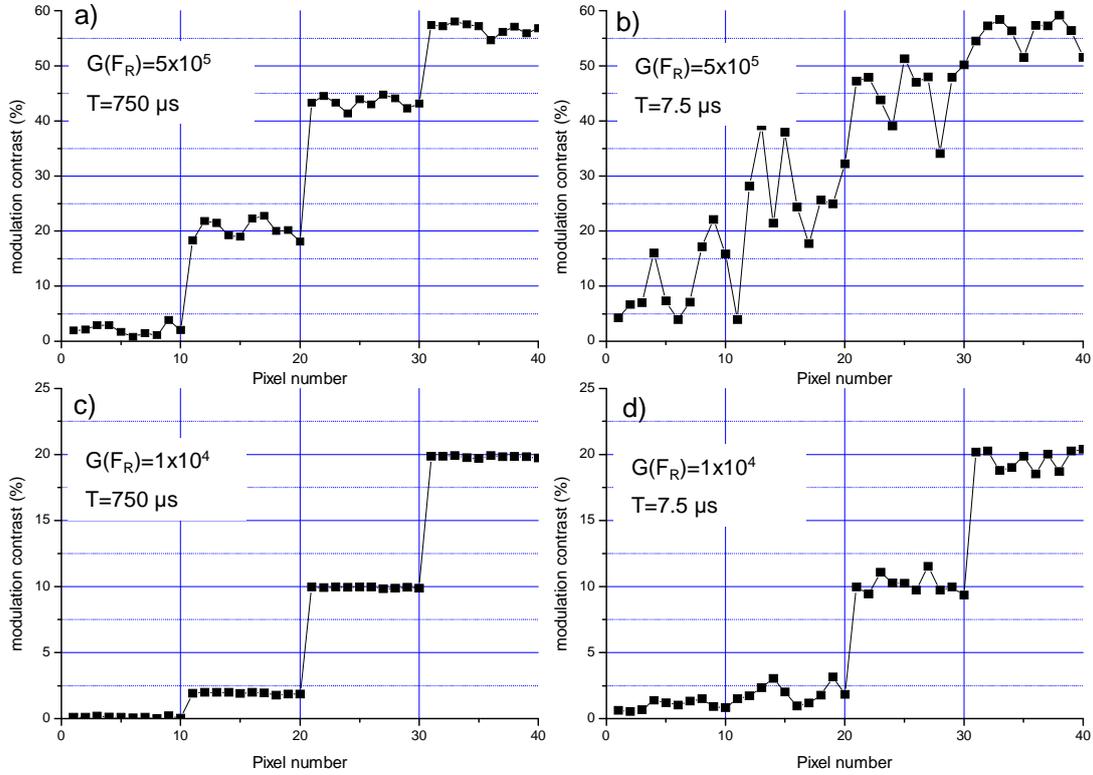

Fig. 4. Numerical 1D scans obtained from the measured laser output power modulation contrast of an autodyne interferometer, when the laser beam is scanned on reflectivity stairs composed of 4 steps. Experimental conditions: Laser output power, $P_{out} = 117\,\text{mW}$ ($p_{out} = 6.25 \times 10^{17}\,\text{photons/s}$ at $\lambda = 1064\,\text{nm}$); Laser relaxation frequency, $F_R = 1.6\,\text{MHz}$; Autodyne modulation frequency, $F_e = F_R$. Step 1: (pixels 1 to 10), $R_{e,1} = 0$; Step 2: (pixels 11 to 20), $R_{e,2} = \dfrac{10^{-10}}{100}$ Step 3: (pixels 21 to 30), $R_{e,3} = \dfrac{10^{-10}}{4}$; Steps 4: (Pixels 31 to 40), $R_{e,4} = 10^{-10}$.

Top row: $G(F_R) = 5 \times 10^5$; Bottom row: $G(F_R) = 1 \times 10^4$; Left column: $\Delta F = 1/(2T) = 666\,\text{Hz}$; Right column): $\Delta F = 1/(2T) = 66.6\,\text{kHz}$.

**Table 1. Modulation contrast (MC) and signal to noise ratio (SNR) of the LOFI images (Figs. 4) obtained with the laser having the lower value LOFI gain $(G(F_R) = 1 \times 10^4)$. Numerical results are in bold while the analytical results are written in italics between parentheses.**

| $R_e$ | 0 | $1 \times 10^{-12}$ | $2.5 \times 10^{-11}$ | $1 \times 10^{-10}$ |
|---|---|---|---|---|
| $\Delta F = 666\,\text{Hz}$ | | | | |
| MC(%) | **0.1** (*0*) | **1.9** (*2.0*) | **9.9** (*9.9*) | **19.8** (*19.9*) |
| SNR | **1** (*1*) | **19.4** (*21.7*) | **100.3** (*108.3*) | **200.6** (*216.5*) |
| $\Delta F = 66.6\,\text{kHz}$ | | | | |
| MC(%) | **1.0** (*0*) | **1.9** (*2.0*) | **10.1** (*9.9*) | **19.6** (*19.9*) |
| SNR | **1** (*1*) | **1.9** (*2.2*) | **10.1** (*10.8*) | **19.6** (*21.7*) |

**Table 2. Modulation contrast (MC) and signal to noise ratio (SNR) of the LOFI images (Figs. 4) obtained with the laser having the higher value LOFI gain ($G(F_R) = 5 \times 10^5$). Numerical results are in bold while the analytical results are written in italics between parentheses.**

| $R_e$ | 0 | $1 \times 10^{-12}$ | $2.5 \times 10^{-11}$ | $1 \times 10^{-10}$ |
|---|---|---|---|---|
| $\Delta F = 666\,\text{Hz}$ <br> MC(%) <br> SNR | **2** (*0*) <br> **1** (*1*) | **20.3** (*100*) <br> **9.8** (*21.7*) | **43.4** (*500*) <br> **21.1** (*108.3*) | **56.8** (*1000*) <br> **27.6** (*216.5*) |
| $\Delta F = 66.6\,\text{kHz}$ <br> MC(%) <br> SNR | **10.7** (*0*) <br> **1** (*1*) | **25.6** (*100*) <br> **2.4** (*2.2*) | **45.6** (*500*) <br> **4.3** (*10.8*) | **56.0** (*1000*) <br> **5.2** (*21.7*) |